\documentclass[prl,twocolumn,twoside,preprintnumbers,superscriptaddress]{revtex4}

\usepackage{amsmath}
\usepackage{graphicx}\usepackage{multirow}

\newcommand{\eq}[1]{Eq.~(\ref{#1})}

\newcommand{\fig}[1]{Fig.~\ref{#1}}

\newcommand{\GeV}{\,\text{GeV}}
\newcommand{\TeV}{\,\text{TeV}}

\newcommand{\SMchiEFT}{${\rm SM}_\chi$ EFT}

\begin{document}

\preprint{\vbox{\hbox{UCB-PTH-14/01}}}
\preprint{\vbox{\hbox{CERN-PH-TH/2014-016}}}

\title{New Constraints on Dark Matter Effective Theories from Standard Model Loops}

\author{Andreas Crivellin}
\affiliation{CERN Theory Division, CH-1211 Geneva 23, Switzerland}%
\affiliation{Albert Einstein Center for Fundamental
Physics, \\ 
Institute for Theoretical Physics, University of Bern, CH-3012
Bern, Switzerland}%
\author{Francesco~D'Eramo}
\affiliation{Department of Physics, University of California, Berkeley, CA 94720, USA}
\affiliation{Theoretical Physics Group, Lawrence Berkeley National Laboratory, Berkeley, CA 94720, USA}
\author{Massimiliano~Procura}
\affiliation{Albert Einstein Center for Fundamental Physics,\\
Institute for Theoretical Physics, University of Bern, Switzerland}%

\date{May 16, 2014}
\begin{abstract}

We consider an effective field theory for a gauge singlet Dirac dark matter (DM) particle interacting with the Standard Model (SM) fields via effective operators suppressed by the scale $\Lambda \gtrsim 1$~TeV. We perform a systematic analysis of the leading loop contributions to spin-independent (SI) DM--nucleon scattering using renormalization group evolution between $\Lambda$ and the low-energy scale probed by direct detection experiments. We find that electroweak interactions induce operator mixings such that operators that are naively velocity-suppressed and spin-dependent can actually contribute to SI scattering. This allows us to put novel constraints on Wilson coefficients that were so far poorly bounded by direct detection. Constraints from current searches are already significantly stronger than LHC bounds, and will improve in the near future. Interestingly, the loop contribution we find is isospin violating even if the underlying theory is isospin conserving. 

 \end{abstract}
%

\maketitle

\noindent 
{\bf Introduction.} A Weakly Interacting Massive Particle (WIMP) is an appealing dark matter (DM) candidate~\cite{Lee:1977ua,Jungman:1995df,Bertone:2004pz,Feng:2010gw}. The lack of evidence for New Physics at the Fermi scale motivates us to remain unbiased about the nature of DM and pursue model-independent approaches. Assuming that the DM is the only non-SM particle experimentally accessible is not always justified at colliders~\cite{Shoemaker:2011vi,Busoni:2013lha,Profumo:2013hqa}, and simplified models have been recently proposed to overcome this limitation~\cite{Chang:2013oia,An:2013xka,DiFranzo:2013vra,Buchmueller:2013dya}. Nevertheless, besides very specific cases (e.g. inelastic DM~\cite{TuckerSmith:2001hy}), it is an excellent approximation for direct searches given the small energy exchanged with the target nuclei. Within this approach, DM interactions with SM fields can be parameterized by higher dimensional operators suppressed by the cutoff scale $\Lambda$, with the main strength of providing model-independent relations among distinct null DM searches~\cite{Goodman:2010yf,Goodman:2010ku,Rajaraman:2011wf,Bai:2012xg,Bai:2010hh,Fox:2011fx,Fox:2011pm,Agrawal:2010fh}. However, different search strategies probe different energy scales, and such a separation of scales may have striking consequences when a connection between different experiments or ultraviolet (UV) complete models with experiments is attempted. Indeed, in some cases loop corrections are known to dramatically alter direct detection (DD) rates~\cite{Kopp:2009et,Freytsis:2010ne,Hill:2011be,Hill:2013hoa,Hill:2014yka,Frandsen:2012db,Haisch:2012kf,Haisch:2013uaa,Kopp:2014tsa}.

In this article we consider the case of a SM gauge singlet Dirac DM ($\chi$), with $m_\chi < \Lambda$, and we calculate the complete set of one-loop effects induced by SM fields for operators up to dimension 6 that contribute to spin-independent (SI) DM--nucleon scattering. The separation between $\Lambda$ and the DD scale is systematically taken into account via a proper renormalization group (RG) analysis. This procedure requires as a first step the computation of both electroweak (EW) and QCD running from the scale $\Lambda$ to the EW symmetry breaking scale where threshold corrections are calculated and the heavy SM fields (Higgs, top quark, $W$ and $Z$ bosons) get integrated out, giving rise to new operators. Then the evolution of these operators from the EW symmetry breaking scale down to the hadronic scale relevant for DD is performed evaluating both QCD and QED contributions. 

We find that while the (known, see e.g.~\cite{Vecchi:2013iza}) QCD effects turn out to be numerically negligible in our case, EW corrections play an important role. In particular, we identify a mixing of a dimension-6 operator whose signal is both spin-dependent (SD) and velocity suppressed into one giving unsuppressed contributions to SI scattering. Our calculation allows us to constrain the former with direct searches. Our limits on the messenger scale $\Lambda$ turns out to improve by an one order of magnitude the collider bounds, and thanks to forthcoming DD experiments will get stronger soon. Our model-independent approach implies that the mixing we find is present in any UV completion which dominantly generates dimension-6 operators with axial quark currents, as for example in $Z^\prime$-portal models~\cite{Alves:2013tqa,Arcadi:2013qia}. In addition, we compute the mixing between DM operators with heavy and light quarks induced by single-photon exchange, which in the case of non-universal DM--quark couplings can be used to probe couplings to heavy quarks.

\vspace{0.1cm}
\noindent 
{\bf SM$_\chi$ EFT.}  Our conceptual starting point is a renormalizable theory of New Physics, where interactions between the DM particle $\chi$ and SM fields are mediated by heavy messenger particles with masses of order $\Lambda \gtrsim 1 \TeV$. Upon integrating out the heavy mediators, the UV complete model is matched at the scale $\Lambda$ onto what we call ``\SMchiEFT", an effective field theory (EFT) whose dynamical degrees of freedom are $\chi$ and the SM fields. The resulting effective Lagrangian has the schematic form
\begin{equation}
{\cal L}_{{\rm SM}_\chi} = \sum_{d > 4}{\cal L}_{{\rm SM}_\chi}^{(d)}\,, \;\;\;\;\;\;\;\; 
{\cal L}_{{\rm SM}_\chi}^{(d)} = \sum_\alpha C_\alpha^{(d)} \, {O}_\alpha^{(d)}~.
\end{equation}
Here, $\alpha$ runs over all possible operators of dimension $d$ allowed by the SM gauge symmetries, which are suppressed by powers of the EFT cutoff scale $\Lambda$ as $1/\Lambda^{d-4}$. The Wilson coefficients $C_\alpha^{(d)}$ are dimensionless, in general scale dependent, and encode unresolved dynamics at higher scales. DM stability forbids operators with just one DM field, and we do not need more than two $\chi$ fields for our study. 
By applying Fierz identities, each operator can be expressed as the product of a DM bilinear and a SM-singlet operator built only with SM fields.
A basis of operators for DD is obtained following the same procedure described in~\cite{Buchmuller:1985jz,Grzadkowski:2010es} for pure SM fields. In what follows, we focus on operators up to dimension~6 generated at the matching scale $\Lambda$, and consistently and systematically derive their effects for DD.

At the matching scale where $SU(2)_L$ is unbroken, four effective operators contribute to DM--nucleon scattering at $d=5$, i.e. the magnetic and electric dipole operators
\begin{align} \label{VAop}
O_{M}^{T} = \frac{1}{{{\Lambda}}}\bar \chi {\sigma ^{\mu\nu} }\chi\, B_{\mu\nu}\,, \qquad O_{E}^{T} = \frac{i}{{{\Lambda}}}\bar \chi {\sigma ^{\mu\nu} \gamma^5} \chi\, B_{\mu\nu}
\end{align}
and the Higgs operators
\begin{align} \label{VAop2}
O_{HH}^S = \frac{{{1 }}}{{{\Lambda}}}\bar \chi \chi  \, H^\dag H  \,,\qquad O_{HH}^P = \frac{{{i }}}{{{\Lambda}}}\bar \chi \gamma^5 \chi  \, H^\dag H \,.
\end{align}
$B_{\mu \nu}$ and $H$ are the $U(1)_Y$ field strength tensor and the SM Higgs doublet, respectively. At $d=6$, tree-level exchange of messengers can generate interactions between DM currents and either quark\footnote{DM interactions in which lepton couplings are dominant are not considered here.} or Higgs currents\footnote{The Wilson coefficients of left-handed up and down quarks must be identical at the scale $\Lambda$ to respect $SU(2)_L$ gauge invariance.} 
\begin{align} \label{eq:currents}
& O_{qq}^{IJ} = \frac{1}{{{\Lambda ^2}}} \, \bar\chi {\Gamma_I^\mu } \chi  \; 
\bar q \, \Gamma_{J,\mu}\,q \,, \notag\\
& O_{HH D}^I = \frac{i}{{{\Lambda ^2}}} \bar \chi \, {\Gamma_I^\mu } \chi \, [{H^\dag }\overleftrightarrow{D}_\mu H\,] \,,
\end{align}
where $q$ runs over the quark flavors, while $I$ and $J$ stand for either $V$ or $A$, with $\Gamma_V^\mu = \gamma^\mu$ and $\Gamma_A^\mu = \gamma^\mu \gamma^5$. We define ${H^\dag }\overleftrightarrow{D}^\mu H \equiv H^\dag ({D}^\mu H) - ({D}^\mu H)^\dag H$. These are all operators at the scale $\Lambda$ up to dimension~6 that can contribute to the SI cross section.\footnote{The dimension-6 interactions between DM currents and $\partial^\mu B_{\mu\nu}$ can be removed by a field redefinition~\cite{Arzt:1993gz}.} We now investigate their effects on the DD rates.

In the effective Lagrangian for elastic WIMP--nucleon scattering, the heavier SM fields (Higgs, $W$ and $Z$ bosons and $t, b, c$ quarks) have to be integrated out and the Higgs' vacuum expectation value gives rise to quark masses. Therefore, among the operators above only $O_{uu, dd}^{VV}$ enter {\it directly} the SI cross section while threshold corrections from the dimension-5 $O^S_{HH}$ generate dimension-7 scalar contributions. The DM--nucleon SI cross section accordingly reads (cf.~\cite{Rajaraman:2011wf,Hill:2011be,Cirigliano:2012pq,Crivellin:2013ipa})
\begin{align}\label{SIeq}
\sigma _N^\text{SI}=& \frac{m_\chi^2\, m_N^2}{(m_\chi + m_N)^2\,\pi\,\Lambda^4}\,\bigg| \sum\limits_{q = u,d} C_{qq}^{VV}f_{V_q}^N
\notag\\
&+
\frac{m_N}{\Lambda}  \bigg(\sum\limits_{q = u,d,s}   C_{qq}^{SS}f_q^N -  12\pi \,C_{gg}^{S}\,f_Q^N  \bigg)  \bigg|^2,
\end{align}
with $m_N$ denoting the nucleon mass, and scalar (vector) couplings $f_q^N$ ($f_{V_q}^N$). For heavy quarks, the parameter $f_Q^N$ is induced by the gluon operator as discussed in~\cite{Shifman}, see also~\cite{Crivellin:2013ipa}. Here, 
\begin{align} \label{qgop}
O_{g g}^S &= \frac{\alpha_s}{{{\Lambda ^3}}}\,\bar{\chi} \chi \,G_{\mu \nu} G^{\mu \nu}\,,\;\;\;
O_{q q}^{SS} = \frac{{{m_q}}}{{{\Lambda ^3}}}\;\bar{\chi} \chi \,\bar{q} q \,,
\end{align}
with $G_{\mu \nu}$ denoting the gluon field strength tensor. In the next section we will discuss how the Wilson coefficients of the operators in \eq{VAop}, \eq{VAop2} and \eq{eq:currents} at the high scale $\Lambda$ are evolved down to the scale of DD and how they are connected to the Wilson coefficients of the low-scale operators in \eq{qgop}.

\vspace{0.1cm}
\noindent 
{\bf Threshold corrections and mixing.} At dimension~5, $O_{M}^{T}$, $O_{E}^{T}$ and $O_{HH}^{S,P}$ do not mix into other operators since they are the lowest dimensional ones, and therefore only threshold corrections have to be computed. The $Z$ boson in $B_{\mu \nu}$, once integrated out, generates $O_{qq}^{VV, VA}$ at dimension~6. The photon field is also encoded in $B_{\mu \nu}$ but it is a degree of freedom of the low-energy theory, and the resulting long-range interaction between $\chi$ and nucleons severely constrains the Wilson coefficient of the dipole operator~\cite{Barger:2010gv,Banks:2010eh,Fortin:2011hv}. The Higgs operator $O_{HH}^S$ gives rise to $O_{q q }^{SS}$ after EW symmetry breaking, and upon integrating out the heavy quarks also the dimension-7 interaction with the gluon field strength $O_{g g}^S$ is generated. This leads to the following threshold corrections
\begin{align} \label{thre_H}
C_{gg}^{S} = \frac{1}{12\pi}\frac{{{\Lambda ^2}}}{{m_{{h^0}}^2}}C_{{HH}}^{S}\,,\qquad C_{qq}^{SS} =  - \frac{\Lambda^2 }{{m_{{h^0}}^2}}C_{HH}^S\, ,
\end{align}
whose form shows that the $O_{qq}^{SS}$ contribution induced by tree-level Higgs exchange is enhanced since it scales like $1/(\Lambda \,m_{{h^0}}^2)$ instead of $1/\Lambda^3$. Typical scattering cross sections involving DM effective couplings to the SM Higgs (like $C_{HH}^S$) are in the ballpark of current experimental limits~\cite{Kim:2006af,Kim:2008pp,MarchRussell:2008yu,Kanemura:2010sh,LopezHonorez:2012kv}. They may also contribute to mono-Higgs production at colliders~\cite{Petrov:2013nia,Carpenter:2013xra}, and for light enough DM ($m_\chi < m_{{h^0}}/2$) to the invisible Higgs decay width~\cite{D'Eramo:2007ga,Pospelov:2011yp,Bai:2011wz,Greljo:2013wja}. 

The evolution matrix for the operators defined in \eq{qgop} only contains one non-vanishing off-diagonal entry, namely $O_{gg}^S$ mixes with $O_{qq}^{SS}$. Using \eq{thre_H}, we find
\begin{align}
&C_{qq}^{SS}\left( \mu_0 \right) = \left[ {\frac{1}{12\pi}\left( {U_{{m_b},{m_t}}^{\left( 5 \right)} + 2\;U_{\mu_0{\rm{,}}{{\rm{m}}_b}}^{\left( 4 \right)}} \right) - 1 } \right]\frac{{{\Lambda ^2}}}{{m_{{h^0}}^2}}C_{{HH}}^{S}  \ ,\nonumber \\
&{\text{with}\;\;\;\;\;}  U_{\mu ,\Lambda }^{\left( n_f \right)} = \frac{{ - 3{C_F}}}{{\pi \beta _0^{}}}\ln {\frac{{{\alpha _s}\left( \Lambda  \right)}}{{{\alpha _s}\left( \mu  \right)}}} \ .
\end{align}
Here, $n_f$ is the number of active flavors, ${\beta _0} = 11 - \frac{2}{3}{n_f}$ and $C_F=4/3$. $\mu_0 < m_b$ is the low-energy scale relevant for DD. The mixing between $O^{SS}_{qq}$ and $O_{gg}^S$ has already been calculated in~\cite{Chetyrkin:1996ke,Hill:2011be,Frandsen:2012db,Vecchi:2013iza}. We find that this has a numerically negligible impact on $\sigma_N^{\text{SI}}$. The reason is that it yields a contribution to $C^{SS}_{qq}$ proportional to $C^{S}_{gg}$ but the effect of $C^{S}_{gg}$ in the cross section is enhanced by a factor of $12 \pi$ compared to the scalar contribution (see \eq{SIeq} and the analysis of the QCD trace anomaly in~\cite{Shifman}).

\begin{figure} 
{\includegraphics[height=2.45cm]{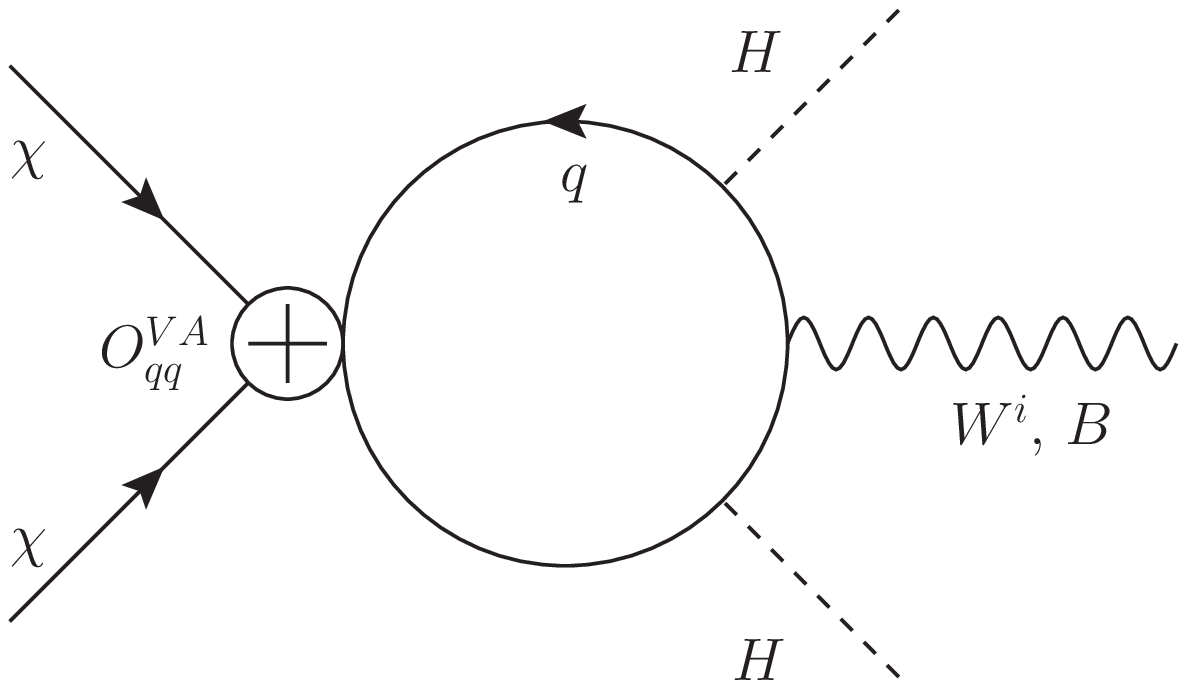}} 
\hfill 
\includegraphics[height=2.45cm]{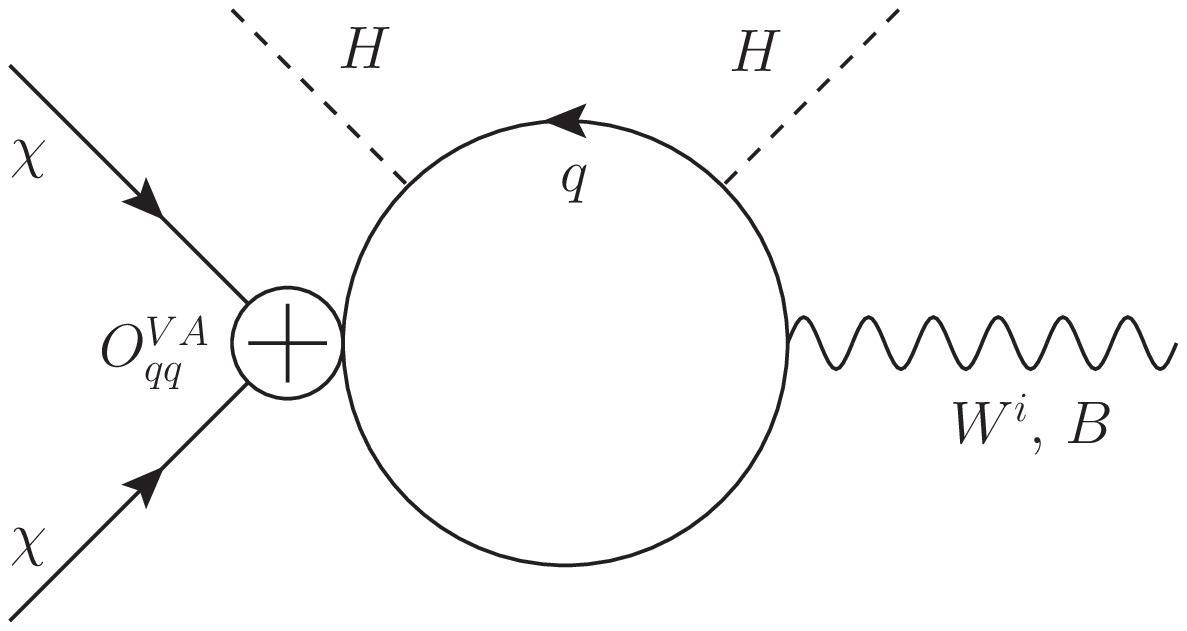}
\vspace{-2ex}
\caption{Diagrams responsible for the mixing of $O^{VA}_{qq}$ into $O_{H H D}^V$. Graphs originated by crossing or reversing the fermion flow are not displayed.}
\vspace{-2ex}
\label{Zgraphs}
\end{figure}

Let us now turn to the dimension-6 operators (see \eq{eq:currents}). Since we focus on SI interactions, only vector DM bilinears are relevant. Concerning quark currents, no QCD renormalization effect has to be taken into account: singlet quark vector currents are conserved under strong interactions and there is no one-loop RG contribution from the axial anomaly. However, EW corrections give rise to an interesting effect which has not been considered so far, namely the mixing of $O_{qq}^{VA}$ into $O_{HH D}^{V}$, which affects DD rates.\footnote{In the DM axial-current sector there is an analogous mixing between $O_{qq}^{AA}$ and $O_{HH D}^A$. Since this only affects SD scattering, it is not relevant for our study. However, for Majorana DM coupling mostly to heavy quarks it can be an important effect.} There are six diagrams contributing to this mixing, two of which are shown in~\fig{Zgraphs}. The result is proportional to the mass of the quark in the loop, i.e. to the Yukawa couplings $Y_q$, and it is therefore dominated by the top quark and to a less extent by the bottom quark. Solving the RG equation, we obtain
\begin{align} \label{VAeq}
C_{HH D}^V\left( \mu  \right) = C_{HH D}^V\left( \Lambda\right) - 
\frac{{{\alpha _t } N_c}}{\pi } C_{tt}^{VA}\left( \Lambda \right) \ln\frac{\mu}{\Lambda} - (t \to b)
\end{align}
with $\alpha_t = Y_t^2/(4 \pi)$. The relative sign between the last two terms is due to the fact that left-handed up- and down-type quarks have opposite eigenvalues of the third weak-isospin component. Here we keep only the top and bottom contributions to the loop. In applying this result, the running scale $\mu$ should be identified with the EW symmetry breaking scale, where the top and the $Z$ are integrated out and the corresponding logarithm is frozen. A non-vanishing value of $C^V_{H H D}$ generates a finite threshold correction to $O^{VV}_{qq}$ and $O^{VA}_{qq}$  at the EW symmetry breaking scale by attaching a quark pair and integrating out the $Z$ boson:
\begin{align}\label{ZTHC}
C_{uu}^{VV} &\to C_{uu}^{VV} \!+\! \left( \frac{1}{2} - \frac{4}{3} s_w^2 \right) C_{HH D}^V~, \\
C_{dd}^{VV} & \to C_{dd}^{VV} \!+\! \left( - \frac{1}{2} + \frac{2}{3} s_w^2 \right) C_{HH D}^V~, \nonumber
\end{align}
where $s_w$ is the sine of the weak mixing angle. Combining  \eq{ZTHC} and \eq{VAeq}, we find
\begin{align}\label{main}
C_{uu}^{VV}\!\left( \mu  \right) &= C_{uu}^{VV}\!\left( \Lambda  \right) + 
\left( \frac{1}{2} - \frac{4}{3} s_w^2 \right) C_{HH D}^V\left( \Lambda\right) + \\ &   
\left( \frac{1}{2} - \frac{4}{3} s_w^2 \right) \left[- \frac{{{\alpha _t } N_c}}{\pi } C_{tt}^{VA}\left( \Lambda \right) \ln\frac{\mu}{\Lambda} - (t \to b) \right] \ , \nonumber \\ 
C_{dd}^{VV}\!\left( \mu  \right) &= C_{dd}^{VV}\!\left( \Lambda  \right) + \left( - \frac{1}{2} + \frac{2}{3} s_w^2 \right) C_{HH D}^V\left( \Lambda \right) + \nonumber \\ &\left( - \frac{1}{2} + \frac{2}{3} s_w^2 \right) \left[- \frac{{{\alpha _t } N_c}}{\pi } C_{tt}^{VA}\left( \Lambda \right) \ln\frac{\mu}{\Lambda} - (t \to b) \right] \ , \nonumber 
\end{align}
which means that a quark vector current is generated at the low scale, even if at the high scale there is only an axial-vector current. As an application of our results, in the next section we will present limits on $C_{qq}^{VA}$, previously bounded only by collider searches (see e.g.~\cite{Zhou:2013fla}). 

We now consider the mixing between vector operators of heavy and light quarks obtained by attaching to the quark loop a photon which couples to a quark pair. Since only $u$- and $d$- quarks contribute to the scattering cross section, this can be used to constrain all quark vector current operators with heavier quarks ($q=s,c,b,t$). This is relevant in the case of non-universal DM-quark couplings. Depending on the number of active flavors, the pertinent RG equation reads
\begin{equation}
\frac{{d\,\vec C(\mu)}}{{d\ln \mu }} = \frac{{{\alpha_{\text{EM}}} N_c}}{{4\pi }}{\gamma_0}\;\vec 
C(\mu),\;\;\;\vec C = \left( {\begin{array}{*{20}{c}}
{\vec C_{dd}^{VV}}\vspace{0.2cm}
\\ 
{\vec C_{uu}^{VV}}
\end{array}} \right) \, ,
\end{equation}
\vspace{-0.4cm}
\begin{equation} 
\renewcommand{\arraystretch}{2.2}
{\gamma _0} = \left( {\begin{array}{*{20}{c}}
{{{\left[ - 8/27 \right]}_{{n_d} \times {n_d}}}}&\vline& {{{\left[
16/27 \right]}_{{n_d} \times {n_u}}}}\\
\hline
{{{\left[ 16/27 \right]}_{{n_u} \times {n_d}}}}&\vline& {{{\left[ -32/27 \right]}_{{n_u} \times {n_u}}}}
\end{array}} \right) \, .
\end{equation}
Here, ${\vec C_{dd}^{VV}}$ and ${\vec C_{uu}^{VV}}$ are vectors in flavor space whose dimension is determined by the number of active flavors, and ${{\left[a\right]}_{{n_d} \times {n_d}}}$ stands for a ${{n_d} \times {n_d}}$ matrix with all entries equal to $a$. A similar mixing induces DM couplings to lepton currents which can play a significant role in structure formation~\cite{Shoemaker:2013tda,Cornell:2013rza}. For DM coupling only to leptons, constraints from DD are induced by loop effects similar to the ones discussed here~\cite{Kopp:2009et,Fox:2011fx,Agrawal:2011ze}.

\begin{figure} 
\includegraphics[height=40ex]{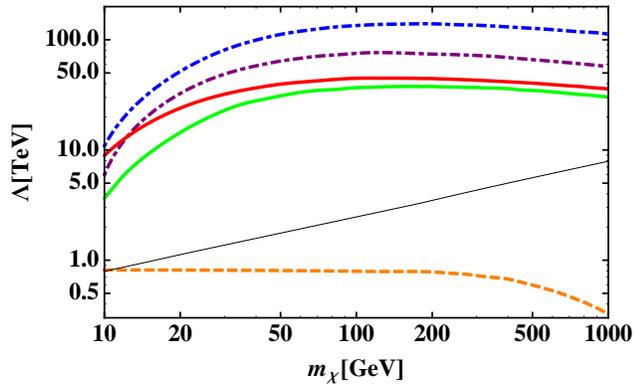}
\caption{Lower bounds for allowed regions from LHC searches~\cite{Zhou:2013fla} (dashed, orange) and SI WIMP--nucleon scattering from XENON100 (solid, green)~\cite{Aprile:2012nq} and LUX (solid, red)~\cite{Akerib:2013tjd}. Projected allowed regions for SCDMS~\cite{SCDMS} (dot-dashed, purple) and XENON1T~\cite{Aprile:2012zx} (dot-dashed, blue) are also shown, as well as the curve giving the correct thermal relic density (thin, black). Here we set $C^{VA}_{qq} = 1$ while all other Wilson coefficients are assumed to be zero.}
\label{plot1}
\end{figure}

\vspace{0.1cm}
\noindent 
{\bf Numerical Analysis.} We use our results to put constraints on Wilson coefficients that have not yet been bounded from direct searches. We first consider the scenario where $C^{VA}_{qq}$ is the only non-vanishing coefficient at the scale $\Lambda$, and assume flavor-universal DM--quark couplings. The regions in parameter space allowed by various experiments are shown in \fig{plot1}, where the matching scale $\Lambda$ is plotted as a function of the DM mass for $C^{VA}_{qq} = 1$.\footnote{In the standard notation of ~\cite{Goodman:2010ku}, our operator $O^{VA}_{qq}$ corresponds to D7 with $M_* = \Lambda/\sqrt{C^{VA}_{qq}}$.} If loop effects are neglected, this operator generates a scattering amplitude which is both SD and velocity suppressed. For this reason, the best bound before our analysis came from collider searches (see e.g.~\cite{Zhou:2013fla}), corresponding to the dashed orange line in \fig{plot1}. The RG induced contribution of $C^{VA}_{qq}$ to $C^{VV}_{dd,uu}$ allows us to equally well constrain this operator from SI measurements. In order to use the experimental bounds on the WIMP--nucleon cross section given in~\cite{Aprile:2012nq,Akerib:2013tjd}, we have to take care of the fact that these limits were obtained under the assumption of negligible isospin violation. However, as we see from \eq{main}, our loop contribution to $C^{VV}_{qq}$ is isospin violating, i.e. $\Delta C^{VV}_{dd} \simeq - 2 \, \Delta C^{VV}_{uu}$.\footnote{We point out the remarkable fact that in our case isospin violation is entirely due to SM loops and is present even if the UV complete theory does not violate isospin.} Therefore, unlike the isospin-symmetric case, our WIMP--nucleus cross section does not scale just like $A^2$ (where $A$ is the mass number of the target nucleus). The regions allowed by DD measurements are delimited by the green (XENON100) and red (LUX) lines. Remarkably, these bounds are one order of magnitude stronger than the ones from LHC searches (dashed orange). We also study the impact of future SI measurements, and show the projections for the allowed regions from SCDMS~\cite{SCDMS} (purple) and XENON1T~\cite{Aprile:2012zx} (blue). We also superimpose the line obtained by requiring a $O^{VA}_{qq}$-dominated thermal freeze-out and observe that current experiments completely rule out the thermal window (for $C^{VA}_{qq} = 1$).

\begin{figure} 
\includegraphics[height=40ex]{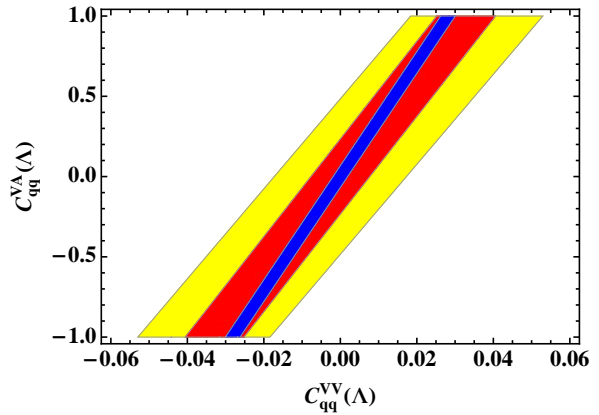}
\caption{Regions in parameter space allowed by LUX~\cite{Akerib:2013tjd} for $\Lambda = 10 \TeV$ (blue), $20 \TeV$ (red) and $30 \TeV$ (yellow), assuming flavor universality in the DM--quark coupling. In the case of non-universal couplings, $C^{VA}_{qq}$ should be replaced by $C^{VA}_{tt}$. The dark matter mass is fixed to $100 \GeV$.}
\label{plot2}
\end{figure}

In the SM$_\chi$ EFT, it is possible to assume that $C^{VA}_{qq}\neq 0$ and $C^{V}_{HHD} = 0$ only at one fixed scale (in \fig{plot1}, this scale is $\Lambda$). We extend our analysis to the case where also $O^{V}_{HHD}$ (and  $O^{VV}_{qq}$) is switched on, and we use the matching corrections in Eq.~(\ref{ZTHC}) to discuss the effect in terms of an effective $C^{VV}_{qq}$ at the matching scale $\Lambda$. In \fig{plot2} we show the parameter space regions allowed by LUX in the $(C^{VV}_{qq}$, $ C^{VA}_{qq})$ plane for different values of $\Lambda$. Any UV complete model generating only (axial-)vector operators must respect these bounds.

\vspace{0.1cm}
\noindent 
{\bf Discussion and Outlook.}  In this article we highlighted the importance of a systematic analysis of one-loop effects induced by SM fields to connect effective operators at the New Physics scale with DD rates. We computed all relevant one-loop effects for SI interactions up to dimension 6 (at the scale $\Lambda$) for a gauge singlet Dirac WIMP. Previously known QCD corrections are numerically not very relevant in this case, although can have drastic effects for electroweak charged candidates (e.g. wino, higgsino~\cite{Hill:2013hoa}). Instead, the new EW corrections that we computed allowed us to use DD data to significantly improve bounds on Wilson coefficients. More specifically, we put constraints on the SD and velocity-suppressed operator $O^{VA}_{qq}$. Our bounds are much stronger than LHC measurements and will significantly improve when new data will become available. For non-universal DM couplings, the mixing we computed between heavy and light quark currents induced by photon exchange allows us to constrain $C^{VV}_{QQ}$ for heavy quarks $Q= s, c, b, t$.

Although an analysis of UV complete models is beyond the scope of this article, we point out that our EW mixing effect can be relevant for $Z^\prime$-portal models~\cite{Alves:2013tqa,Arcadi:2013qia}, if the quarks couple to the $Z^\prime$ only through the axial current (as in some $E_6$ GUT models~\cite{Hewett:1988xc,Langacker:2008yv}). Kinetic and/or mass mixing between the $Z$ and $Z'$ will generate a contribution to $C^{VV}_{qq}$ which is likely to be small compared to $C^{VA}_{qq}$, and has to obey the constraints in \fig{plot2}. 

Our analysis systematically accounts for contributions from operators up to dimension 6 at the scale $\Lambda$. At dimension 7, an important EW mixing effect is already known: the tensor operator $O_{qq}^{TT} = \frac{1}{{{\Lambda ^3}}}\bar \chi {\sigma ^{\mu\nu} }\chi \,\bar q\, H \, \sigma_{\mu\nu}q$ mixes into the dimension-5 dipole operators $O_{M}^{T}$, $O_{E}^{T}$~\cite{Haisch:2013uaa} and the predictions for SI DD rates get sizably affected. This motivates a systematic analysis of all one-loop effects at dimension 7 including EW corrections, building upon the work presented in this article.


\medskip
{\it Acknowledgments.}--- We acknowledge useful conversations with T. Cohen, U. Haisch, R. Hill, M. Hoferichter, E. Mereghetti, Y. Nomura, W. Shepherd, S. Shirai, M. Solon, T. Tait, J. Thaler and L. Ubaldi. F.D. thanks the Aspen Center for Physics for hospitality when this work was initiated. A.C. is supported by a Marie Curie Intra-European Fellowship of the European Community's 7th Framework Programme under
contract number (PIEF-GA-2012-326948). A.C. and M.P. are supported by the Swiss National Science Foundation and by the ``Innovations- und Kooperationsprojekt C-13" of the Schweizerische Universit{\"a}tskonferenz SUK/CRUS. F.D. is supported by the Miller Institute for Basic Research in Science.

\bibliography{BIB}

\begin{thebibliography}{63}
\expandafter\ifx\csname natexlab\endcsname\relax\def\natexlab#1{#1}\fi
\expandafter\ifx\csname bibnamefont\endcsname\relax
  \def\bibnamefont#1{#1}\fi
\expandafter\ifx\csname bibfnamefont\endcsname\relax
  \def\bibfnamefont#1{#1}\fi
\expandafter\ifx\csname citenamefont\endcsname\relax
  \def\citenamefont#1{#1}\fi
\expandafter\ifx\csname url\endcsname\relax
  \def\url#1{\texttt{#1}}\fi
\expandafter\ifx\csname urlprefix\endcsname\relax\def\urlprefix{URL }\fi
\providecommand{\bibinfo}[2]{#2}
\providecommand{\eprint}[2][]{\url{#2}}

\bibitem[{\citenamefont{Lee and Weinberg}(1977)}]{Lee:1977ua}
\bibinfo{author}{\bibfnamefont{B.~W.} \bibnamefont{Lee}} \bibnamefont{and}
  \bibinfo{author}{\bibfnamefont{S.}~\bibnamefont{Weinberg}},
  \bibinfo{journal}{Phys.Rev.Lett.} \textbf{\bibinfo{volume}{39}},
  \bibinfo{pages}{165} (\bibinfo{year}{1977}).

\bibitem[{\citenamefont{Jungman et~al.}(1996)\citenamefont{Jungman,
  Kamionkowski, and Griest}}]{Jungman:1995df}
\bibinfo{author}{\bibfnamefont{G.}~\bibnamefont{Jungman}},
  \bibinfo{author}{\bibfnamefont{M.}~\bibnamefont{Kamionkowski}},
  \bibnamefont{and} \bibinfo{author}{\bibfnamefont{K.}~\bibnamefont{Griest}},
  \bibinfo{journal}{Phys.Rept.} \textbf{\bibinfo{volume}{267}},
  \bibinfo{pages}{195} (\bibinfo{year}{1996}), \eprint{hep-ph/9506380}.

\bibitem[{\citenamefont{Bertone et~al.}(2005)\citenamefont{Bertone, Hooper, and
  Silk}}]{Bertone:2004pz}
\bibinfo{author}{\bibfnamefont{G.}~\bibnamefont{Bertone}},
  \bibinfo{author}{\bibfnamefont{D.}~\bibnamefont{Hooper}}, \bibnamefont{and}
  \bibinfo{author}{\bibfnamefont{J.}~\bibnamefont{Silk}},
  \bibinfo{journal}{Phys.Rept.} \textbf{\bibinfo{volume}{405}},
  \bibinfo{pages}{279} (\bibinfo{year}{2005}), \eprint{hep-ph/0404175}.

\bibitem[{\citenamefont{Feng}(2010)}]{Feng:2010gw}
\bibinfo{author}{\bibfnamefont{J.~L.} \bibnamefont{Feng}},
  \bibinfo{journal}{Ann.Rev.Astron.Astrophys.} \textbf{\bibinfo{volume}{48}},
  \bibinfo{pages}{495} (\bibinfo{year}{2010}), \eprint{1003.0904}.

\bibitem[{\citenamefont{Shoemaker and Vecchi}(2012)}]{Shoemaker:2011vi}
\bibinfo{author}{\bibfnamefont{I.~M.} \bibnamefont{Shoemaker}}
  \bibnamefont{and} \bibinfo{author}{\bibfnamefont{L.}~\bibnamefont{Vecchi}},
  \bibinfo{journal}{Phys.Rev.} \textbf{\bibinfo{volume}{D86}},
  \bibinfo{pages}{015023} (\bibinfo{year}{2012}), \eprint{1112.5457}.

\bibitem[{\citenamefont{Busoni et~al.}(2013)\citenamefont{Busoni, De~Simone,
  Morgante, and Riotto}}]{Busoni:2013lha}
\bibinfo{author}{\bibfnamefont{G.}~\bibnamefont{Busoni}},
  \bibinfo{author}{\bibfnamefont{A.}~\bibnamefont{De~Simone}},
  \bibinfo{author}{\bibfnamefont{E.}~\bibnamefont{Morgante}}, \bibnamefont{and}
  \bibinfo{author}{\bibfnamefont{A.}~\bibnamefont{Riotto}}
  (\bibinfo{year}{2013}), \eprint{1307.2253}.

\bibitem[{\citenamefont{Profumo et~al.}(2013)\citenamefont{Profumo, Shepherd,
  and Tait}}]{Profumo:2013hqa}
\bibinfo{author}{\bibfnamefont{S.}~\bibnamefont{Profumo}},
  \bibinfo{author}{\bibfnamefont{W.}~\bibnamefont{Shepherd}}, \bibnamefont{and}
  \bibinfo{author}{\bibfnamefont{T.}~\bibnamefont{Tait}}
  (\bibinfo{year}{2013}), \eprint{1307.6277}.

\bibitem[{\citenamefont{Chang et~al.}(2013)\citenamefont{Chang, Edezhath,
  Hutchinson, and Luty}}]{Chang:2013oia}
\bibinfo{author}{\bibfnamefont{S.}~\bibnamefont{Chang}},
  \bibinfo{author}{\bibfnamefont{R.}~\bibnamefont{Edezhath}},
  \bibinfo{author}{\bibfnamefont{J.}~\bibnamefont{Hutchinson}},
  \bibnamefont{and} \bibinfo{author}{\bibfnamefont{M.}~\bibnamefont{Luty}}
  (\bibinfo{year}{2013}), \eprint{1307.8120}.

\bibitem[{\citenamefont{An et~al.}(2013)\citenamefont{An, Wang, and
  Zhang}}]{An:2013xka}
\bibinfo{author}{\bibfnamefont{H.}~\bibnamefont{An}},
  \bibinfo{author}{\bibfnamefont{L.-T.} \bibnamefont{Wang}}, \bibnamefont{and}
  \bibinfo{author}{\bibfnamefont{H.}~\bibnamefont{Zhang}}
  (\bibinfo{year}{2013}), \eprint{1308.0592}.

\bibitem[{\citenamefont{DiFranzo et~al.}(2013)\citenamefont{DiFranzo, Nagao,
  Rajaraman, and Tait}}]{DiFranzo:2013vra}
\bibinfo{author}{\bibfnamefont{A.}~\bibnamefont{DiFranzo}},
  \bibinfo{author}{\bibfnamefont{K.~I.} \bibnamefont{Nagao}},
  \bibinfo{author}{\bibfnamefont{A.}~\bibnamefont{Rajaraman}},
  \bibnamefont{and} \bibinfo{author}{\bibfnamefont{T.~M.~P.}
  \bibnamefont{Tait}}, \bibinfo{journal}{JHEP} \textbf{\bibinfo{volume}{1311}},
  \bibinfo{pages}{014} (\bibinfo{year}{2013}), \eprint{1308.2679}.

\bibitem[{\citenamefont{Buchmueller et~al.}(2013)\citenamefont{Buchmueller,
  Dolan, and McCabe}}]{Buchmueller:2013dya}
\bibinfo{author}{\bibfnamefont{O.}~\bibnamefont{Buchmueller}},
  \bibinfo{author}{\bibfnamefont{M.~J.} \bibnamefont{Dolan}}, \bibnamefont{and}
  \bibinfo{author}{\bibfnamefont{C.}~\bibnamefont{McCabe}}
  (\bibinfo{year}{2013}), \eprint{1308.6799}.

\bibitem[{\citenamefont{Tucker-Smith and Weiner}(2001)}]{TuckerSmith:2001hy}
\bibinfo{author}{\bibfnamefont{D.}~\bibnamefont{Tucker-Smith}}
  \bibnamefont{and} \bibinfo{author}{\bibfnamefont{N.}~\bibnamefont{Weiner}},
  \bibinfo{journal}{Phys.Rev.} \textbf{\bibinfo{volume}{D64}},
  \bibinfo{pages}{043502} (\bibinfo{year}{2001}), \eprint{hep-ph/0101138}.

\bibitem[{\citenamefont{Goodman et~al.}(2011)\citenamefont{Goodman, Ibe,
  Rajaraman, Shepherd, Tait et~al.}}]{Goodman:2010yf}
\bibinfo{author}{\bibfnamefont{J.}~\bibnamefont{Goodman}},
  \bibinfo{author}{\bibfnamefont{M.}~\bibnamefont{Ibe}},
  \bibinfo{author}{\bibfnamefont{A.}~\bibnamefont{Rajaraman}},
  \bibinfo{author}{\bibfnamefont{W.}~\bibnamefont{Shepherd}},
  \bibinfo{author}{\bibfnamefont{T.~M.} \bibnamefont{Tait}},
  \bibnamefont{et~al.}, \bibinfo{journal}{Phys.Lett.}
  \textbf{\bibinfo{volume}{B695}}, \bibinfo{pages}{185} (\bibinfo{year}{2011}),
  \eprint{1005.1286}.

\bibitem[{\citenamefont{Goodman et~al.}(2010)\citenamefont{Goodman, Ibe,
  Rajaraman, Shepherd, Tait et~al.}}]{Goodman:2010ku}
\bibinfo{author}{\bibfnamefont{J.}~\bibnamefont{Goodman}},
  \bibinfo{author}{\bibfnamefont{M.}~\bibnamefont{Ibe}},
  \bibinfo{author}{\bibfnamefont{A.}~\bibnamefont{Rajaraman}},
  \bibinfo{author}{\bibfnamefont{W.}~\bibnamefont{Shepherd}},
  \bibinfo{author}{\bibfnamefont{T.~M.} \bibnamefont{Tait}},
  \bibnamefont{et~al.}, \bibinfo{journal}{Phys.Rev.}
  \textbf{\bibinfo{volume}{D82}}, \bibinfo{pages}{116010}
  (\bibinfo{year}{2010}), \eprint{1008.1783}.

\bibitem[{\citenamefont{Rajaraman et~al.}(2011)\citenamefont{Rajaraman,
  Shepherd, Tait, and Wijangco}}]{Rajaraman:2011wf}
\bibinfo{author}{\bibfnamefont{A.}~\bibnamefont{Rajaraman}},
  \bibinfo{author}{\bibfnamefont{W.}~\bibnamefont{Shepherd}},
  \bibinfo{author}{\bibfnamefont{T.~M.} \bibnamefont{Tait}}, \bibnamefont{and}
  \bibinfo{author}{\bibfnamefont{A.~M.} \bibnamefont{Wijangco}},
  \bibinfo{journal}{Phys.Rev.} \textbf{\bibinfo{volume}{D84}},
  \bibinfo{pages}{095013} (\bibinfo{year}{2011}), \eprint{1108.1196}.

\bibitem[{\citenamefont{Bai and Tait}(2013)}]{Bai:2012xg}
\bibinfo{author}{\bibfnamefont{Y.}~\bibnamefont{Bai}} \bibnamefont{and}
  \bibinfo{author}{\bibfnamefont{T.~M.} \bibnamefont{Tait}},
  \bibinfo{journal}{Phys.Lett.} \textbf{\bibinfo{volume}{B723}},
  \bibinfo{pages}{384} (\bibinfo{year}{2013}), \eprint{1208.4361}.

\bibitem[{\citenamefont{Bai et~al.}(2010)\citenamefont{Bai, Fox, and
  Harnik}}]{Bai:2010hh}
\bibinfo{author}{\bibfnamefont{Y.}~\bibnamefont{Bai}},
  \bibinfo{author}{\bibfnamefont{P.~J.} \bibnamefont{Fox}}, \bibnamefont{and}
  \bibinfo{author}{\bibfnamefont{R.}~\bibnamefont{Harnik}},
  \bibinfo{journal}{JHEP} \textbf{\bibinfo{volume}{1012}}, \bibinfo{pages}{048}
  (\bibinfo{year}{2010}), \eprint{1005.3797}.

\bibitem[{\citenamefont{Fox et~al.}(2011)\citenamefont{Fox, Harnik, Kopp, and
  Tsai}}]{Fox:2011fx}
\bibinfo{author}{\bibfnamefont{P.~J.} \bibnamefont{Fox}},
  \bibinfo{author}{\bibfnamefont{R.}~\bibnamefont{Harnik}},
  \bibinfo{author}{\bibfnamefont{J.}~\bibnamefont{Kopp}}, \bibnamefont{and}
  \bibinfo{author}{\bibfnamefont{Y.}~\bibnamefont{Tsai}},
  \bibinfo{journal}{Phys.Rev.} \textbf{\bibinfo{volume}{D84}},
  \bibinfo{pages}{014028} (\bibinfo{year}{2011}), \eprint{1103.0240}.

\bibitem[{\citenamefont{Fox et~al.}(2012)\citenamefont{Fox, Harnik, Kopp, and
  Tsai}}]{Fox:2011pm}
\bibinfo{author}{\bibfnamefont{P.~J.} \bibnamefont{Fox}},
  \bibinfo{author}{\bibfnamefont{R.}~\bibnamefont{Harnik}},
  \bibinfo{author}{\bibfnamefont{J.}~\bibnamefont{Kopp}}, \bibnamefont{and}
  \bibinfo{author}{\bibfnamefont{Y.}~\bibnamefont{Tsai}},
  \bibinfo{journal}{Phys.Rev.} \textbf{\bibinfo{volume}{D85}},
  \bibinfo{pages}{056011} (\bibinfo{year}{2012}), \eprint{1109.4398}.

\bibitem[{\citenamefont{Agrawal et~al.}(2010)\citenamefont{Agrawal, Chacko,
  Kilic, and Mishra}}]{Agrawal:2010fh}
\bibinfo{author}{\bibfnamefont{P.}~\bibnamefont{Agrawal}},
  \bibinfo{author}{\bibfnamefont{Z.}~\bibnamefont{Chacko}},
  \bibinfo{author}{\bibfnamefont{C.}~\bibnamefont{Kilic}}, \bibnamefont{and}
  \bibinfo{author}{\bibfnamefont{R.~K.} \bibnamefont{Mishra}}
  (\bibinfo{year}{2010}), \eprint{1003.1912}.

\bibitem[{\citenamefont{Kopp et~al.}(2009)\citenamefont{Kopp, Niro, Schwetz,
  and Zupan}}]{Kopp:2009et}
\bibinfo{author}{\bibfnamefont{J.}~\bibnamefont{Kopp}},
  \bibinfo{author}{\bibfnamefont{V.}~\bibnamefont{Niro}},
  \bibinfo{author}{\bibfnamefont{T.}~\bibnamefont{Schwetz}}, \bibnamefont{and}
  \bibinfo{author}{\bibfnamefont{J.}~\bibnamefont{Zupan}},
  \bibinfo{journal}{Phys.Rev.} \textbf{\bibinfo{volume}{D80}},
  \bibinfo{pages}{083502} (\bibinfo{year}{2009}), \eprint{0907.3159}.

\bibitem[{\citenamefont{Freytsis and Ligeti}(2011)}]{Freytsis:2010ne}
\bibinfo{author}{\bibfnamefont{M.}~\bibnamefont{Freytsis}} \bibnamefont{and}
  \bibinfo{author}{\bibfnamefont{Z.}~\bibnamefont{Ligeti}},
  \bibinfo{journal}{Phys.Rev.} \textbf{\bibinfo{volume}{D83}},
  \bibinfo{pages}{115009} (\bibinfo{year}{2011}), \eprint{1012.5317}.

\bibitem[{\citenamefont{Hill and Solon}(2012)}]{Hill:2011be}
\bibinfo{author}{\bibfnamefont{R.~J.} \bibnamefont{Hill}} \bibnamefont{and}
  \bibinfo{author}{\bibfnamefont{M.~P.} \bibnamefont{Solon}},
  \bibinfo{journal}{Phys.Lett.} \textbf{\bibinfo{volume}{B707}},
  \bibinfo{pages}{539} (\bibinfo{year}{2012}), \eprint{1111.0016}.

\bibitem[{\citenamefont{Hill and Solon}(2013)}]{Hill:2013hoa}
\bibinfo{author}{\bibfnamefont{R.~J.} \bibnamefont{Hill}} \bibnamefont{and}
  \bibinfo{author}{\bibfnamefont{M.~P.} \bibnamefont{Solon}}
  (\bibinfo{year}{2013}), \eprint{1309.4092}.

\bibitem[{\citenamefont{Hill and Solon}(2014)}]{Hill:2014yka}
\bibinfo{author}{\bibfnamefont{R.~J.} \bibnamefont{Hill}} \bibnamefont{and}
  \bibinfo{author}{\bibfnamefont{M.~P.} \bibnamefont{Solon}}
  (\bibinfo{year}{2014}), \eprint{1401.3339}.

\bibitem[{\citenamefont{Frandsen et~al.}(2012)\citenamefont{Frandsen, Haisch,
  Kahlhoefer, Mertsch, and Schmidt-Hoberg}}]{Frandsen:2012db}
\bibinfo{author}{\bibfnamefont{M.~T.} \bibnamefont{Frandsen}},
  \bibinfo{author}{\bibfnamefont{U.}~\bibnamefont{Haisch}},
  \bibinfo{author}{\bibfnamefont{F.}~\bibnamefont{Kahlhoefer}},
  \bibinfo{author}{\bibfnamefont{P.}~\bibnamefont{Mertsch}}, \bibnamefont{and}
  \bibinfo{author}{\bibfnamefont{K.}~\bibnamefont{Schmidt-Hoberg}},
  \bibinfo{journal}{JCAP} \textbf{\bibinfo{volume}{1210}}, \bibinfo{pages}{033}
  (\bibinfo{year}{2012}), \eprint{1207.3971}.

\bibitem[{\citenamefont{Haisch et~al.}(2013)\citenamefont{Haisch, Kahlhoefer,
  and Unwin}}]{Haisch:2012kf}
\bibinfo{author}{\bibfnamefont{U.}~\bibnamefont{Haisch}},
  \bibinfo{author}{\bibfnamefont{F.}~\bibnamefont{Kahlhoefer}},
  \bibnamefont{and} \bibinfo{author}{\bibfnamefont{J.}~\bibnamefont{Unwin}},
  \bibinfo{journal}{JHEP} \textbf{\bibinfo{volume}{1307}}, \bibinfo{pages}{125}
  (\bibinfo{year}{2013}), \eprint{1208.4605}.

\bibitem[{\citenamefont{Haisch and Kahlhoefer}(2013)}]{Haisch:2013uaa}
\bibinfo{author}{\bibfnamefont{U.}~\bibnamefont{Haisch}} \bibnamefont{and}
  \bibinfo{author}{\bibfnamefont{F.}~\bibnamefont{Kahlhoefer}},
  \bibinfo{journal}{JCAP} \textbf{\bibinfo{volume}{1304}}, \bibinfo{pages}{050}
  (\bibinfo{year}{2013}), \eprint{1302.4454}.

\bibitem[{\citenamefont{Kopp et~al.}(2014)\citenamefont{Kopp, Michaels, and
  Smirnov}}]{Kopp:2014tsa}
\bibinfo{author}{\bibfnamefont{J.}~\bibnamefont{Kopp}},
  \bibinfo{author}{\bibfnamefont{L.}~\bibnamefont{Michaels}}, \bibnamefont{and}
  \bibinfo{author}{\bibfnamefont{J.}~\bibnamefont{Smirnov}}
  (\bibinfo{year}{2014}), \eprint{1401.6457}.

\bibitem[{\citenamefont{Vecchi}(2013)}]{Vecchi:2013iza}
\bibinfo{author}{\bibfnamefont{L.}~\bibnamefont{Vecchi}}
  (\bibinfo{year}{2013}), \eprint{1312.5695}.

\bibitem[{\citenamefont{Alves et~al.}(2013)\citenamefont{Alves, Profumo, and
  Queiroz}}]{Alves:2013tqa}
\bibinfo{author}{\bibfnamefont{A.}~\bibnamefont{Alves}},
  \bibinfo{author}{\bibfnamefont{S.}~\bibnamefont{Profumo}}, \bibnamefont{and}
  \bibinfo{author}{\bibfnamefont{F.~S.} \bibnamefont{Queiroz}}
  (\bibinfo{year}{2013}), \eprint{1312.5281}.

\bibitem[{\citenamefont{Arcadi et~al.}(2013)\citenamefont{Arcadi, Mambrini,
  Tytgat, and Zaldivar}}]{Arcadi:2013qia}
\bibinfo{author}{\bibfnamefont{G.}~\bibnamefont{Arcadi}},
  \bibinfo{author}{\bibfnamefont{Y.}~\bibnamefont{Mambrini}},
  \bibinfo{author}{\bibfnamefont{M.~H.~G.} \bibnamefont{Tytgat}},
  \bibnamefont{and} \bibinfo{author}{\bibfnamefont{B.}~\bibnamefont{Zaldivar}}
  (\bibinfo{year}{2013}), \eprint{1401.0221}.

\bibitem[{\citenamefont{Buchm{\"u}ller and Wyler}(1986)}]{Buchmuller:1985jz}
\bibinfo{author}{\bibfnamefont{W.}~\bibnamefont{Buchm{\"u}ller}}
  \bibnamefont{and} \bibinfo{author}{\bibfnamefont{D.}~\bibnamefont{Wyler}},
  \bibinfo{journal}{Nucl.Phys.} \textbf{\bibinfo{volume}{B268}},
  \bibinfo{pages}{621} (\bibinfo{year}{1986}).

\bibitem[{\citenamefont{Grzadkowski et~al.}(2010)\citenamefont{Grzadkowski,
  Iskrzynski, Misiak, and Rosiek}}]{Grzadkowski:2010es}
\bibinfo{author}{\bibfnamefont{B.}~\bibnamefont{Grzadkowski}},
  \bibinfo{author}{\bibfnamefont{M.}~\bibnamefont{Iskrzynski}},
  \bibinfo{author}{\bibfnamefont{M.}~\bibnamefont{Misiak}}, \bibnamefont{and}
  \bibinfo{author}{\bibfnamefont{J.}~\bibnamefont{Rosiek}},
  \bibinfo{journal}{JHEP} \textbf{\bibinfo{volume}{1010}}, \bibinfo{pages}{085}
  (\bibinfo{year}{2010}), \eprint{1008.4884}.

\bibitem[{\citenamefont{Cirigliano et~al.}(2012)\citenamefont{Cirigliano,
  Graesser, and Ovanesyan}}]{Cirigliano:2012pq}
\bibinfo{author}{\bibfnamefont{V.}~\bibnamefont{Cirigliano}},
  \bibinfo{author}{\bibfnamefont{M.~L.} \bibnamefont{Graesser}},
  \bibnamefont{and}
  \bibinfo{author}{\bibfnamefont{G.}~\bibnamefont{Ovanesyan}},
  \bibinfo{journal}{JHEP} \textbf{\bibinfo{volume}{1210}}, \bibinfo{pages}{025}
  (\bibinfo{year}{2012}), \eprint{1205.2695}.

\bibitem[{\citenamefont{Crivellin et~al.}(2013)\citenamefont{Crivellin,
  Hoferichter, and Procura}}]{Crivellin:2013ipa}
\bibinfo{author}{\bibfnamefont{A.}~\bibnamefont{Crivellin}},
  \bibinfo{author}{\bibfnamefont{M.}~\bibnamefont{Hoferichter}},
  \bibnamefont{and} \bibinfo{author}{\bibfnamefont{M.}~\bibnamefont{Procura}}
  (\bibinfo{year}{2013}), \eprint{1312.4951}.

\bibitem[{\citenamefont{Shifman et~al.}(1978)\citenamefont{Shifman, Vainshtein,
  and Zakharov}}]{Shifman}
\bibinfo{author}{\bibfnamefont{M.~A.} \bibnamefont{Shifman}},
  \bibinfo{author}{\bibfnamefont{A.}~\bibnamefont{Vainshtein}},
  \bibnamefont{and} \bibinfo{author}{\bibfnamefont{V.~I.}
  \bibnamefont{Zakharov}}, \bibinfo{journal}{Phys.Lett.}
  \textbf{\bibinfo{volume}{B78}}, \bibinfo{pages}{443} (\bibinfo{year}{1978}).

\bibitem[{\citenamefont{Barger et~al.}(2011)\citenamefont{Barger, Keung, and
  Marfatia}}]{Barger:2010gv}
\bibinfo{author}{\bibfnamefont{V.}~\bibnamefont{Barger}},
  \bibinfo{author}{\bibfnamefont{W.-Y.} \bibnamefont{Keung}}, \bibnamefont{and}
  \bibinfo{author}{\bibfnamefont{D.}~\bibnamefont{Marfatia}},
  \bibinfo{journal}{Phys.Lett.} \textbf{\bibinfo{volume}{B696}},
  \bibinfo{pages}{74} (\bibinfo{year}{2011}), \eprint{1007.4345}.

\bibitem[{\citenamefont{Banks et~al.}(2010)\citenamefont{Banks, Fortin, and
  Thomas}}]{Banks:2010eh}
\bibinfo{author}{\bibfnamefont{T.}~\bibnamefont{Banks}},
  \bibinfo{author}{\bibfnamefont{J.-F.} \bibnamefont{Fortin}},
  \bibnamefont{and} \bibinfo{author}{\bibfnamefont{S.}~\bibnamefont{Thomas}}
  (\bibinfo{year}{2010}), \eprint{1007.5515}.

\bibitem[{\citenamefont{Fortin and Tait}(2012)}]{Fortin:2011hv}
\bibinfo{author}{\bibfnamefont{J.-F.} \bibnamefont{Fortin}} \bibnamefont{and}
  \bibinfo{author}{\bibfnamefont{T.~M.} \bibnamefont{Tait}},
  \bibinfo{journal}{Phys.Rev.} \textbf{\bibinfo{volume}{D85}},
  \bibinfo{pages}{063506} (\bibinfo{year}{2012}), \eprint{1103.3289}.

\bibitem[{\citenamefont{Kim and Lee}(2007)}]{Kim:2006af}
\bibinfo{author}{\bibfnamefont{Y.~G.} \bibnamefont{Kim}} \bibnamefont{and}
  \bibinfo{author}{\bibfnamefont{K.~Y.} \bibnamefont{Lee}},
  \bibinfo{journal}{Phys.Rev.} \textbf{\bibinfo{volume}{D75}},
  \bibinfo{pages}{115012} (\bibinfo{year}{2007}), \eprint{hep-ph/0611069}.

\bibitem[{\citenamefont{Kim et~al.}(2008)\citenamefont{Kim, Lee, and
  Shin}}]{Kim:2008pp}
\bibinfo{author}{\bibfnamefont{Y.~G.} \bibnamefont{Kim}},
  \bibinfo{author}{\bibfnamefont{K.~Y.} \bibnamefont{Lee}}, \bibnamefont{and}
  \bibinfo{author}{\bibfnamefont{S.}~\bibnamefont{Shin}},
  \bibinfo{journal}{JHEP} \textbf{\bibinfo{volume}{0805}}, \bibinfo{pages}{100}
  (\bibinfo{year}{2008}), \eprint{0803.2932}.

\bibitem[{\citenamefont{March-Russell et~al.}(2008)\citenamefont{March-Russell,
  West, Cumberbatch, and Hooper}}]{MarchRussell:2008yu}
\bibinfo{author}{\bibfnamefont{J.}~\bibnamefont{March-Russell}},
  \bibinfo{author}{\bibfnamefont{S.~M.} \bibnamefont{West}},
  \bibinfo{author}{\bibfnamefont{D.}~\bibnamefont{Cumberbatch}},
  \bibnamefont{and} \bibinfo{author}{\bibfnamefont{D.}~\bibnamefont{Hooper}},
  \bibinfo{journal}{JHEP} \textbf{\bibinfo{volume}{0807}}, \bibinfo{pages}{058}
  (\bibinfo{year}{2008}), \eprint{0801.3440}.

\bibitem[{\citenamefont{Kanemura et~al.}(2010)\citenamefont{Kanemura,
  Matsumoto, Nabeshima, and Okada}}]{Kanemura:2010sh}
\bibinfo{author}{\bibfnamefont{S.}~\bibnamefont{Kanemura}},
  \bibinfo{author}{\bibfnamefont{S.}~\bibnamefont{Matsumoto}},
  \bibinfo{author}{\bibfnamefont{T.}~\bibnamefont{Nabeshima}},
  \bibnamefont{and} \bibinfo{author}{\bibfnamefont{N.}~\bibnamefont{Okada}},
  \bibinfo{journal}{Phys.Rev.} \textbf{\bibinfo{volume}{D82}},
  \bibinfo{pages}{055026} (\bibinfo{year}{2010}), \eprint{1005.5651}.

\bibitem[{\citenamefont{Lopez-Honorez et~al.}(2012)\citenamefont{Lopez-Honorez,
  Schwetz, and Zupan}}]{LopezHonorez:2012kv}
\bibinfo{author}{\bibfnamefont{L.}~\bibnamefont{Lopez-Honorez}},
  \bibinfo{author}{\bibfnamefont{T.}~\bibnamefont{Schwetz}}, \bibnamefont{and}
  \bibinfo{author}{\bibfnamefont{J.}~\bibnamefont{Zupan}},
  \bibinfo{journal}{Phys.Lett.} \textbf{\bibinfo{volume}{B716}},
  \bibinfo{pages}{179} (\bibinfo{year}{2012}), \eprint{1203.2064}.

\bibitem[{\citenamefont{Petrov and Shepherd}(2013)}]{Petrov:2013nia}
\bibinfo{author}{\bibfnamefont{A.~A.} \bibnamefont{Petrov}} \bibnamefont{and}
  \bibinfo{author}{\bibfnamefont{W.}~\bibnamefont{Shepherd}}
  (\bibinfo{year}{2013}), \eprint{1311.1511}.

\bibitem[{\citenamefont{Carpenter et~al.}(2013)\citenamefont{Carpenter,
  DiFranzo, Mulhearn, Shimmin, Tulin et~al.}}]{Carpenter:2013xra}
\bibinfo{author}{\bibfnamefont{L.}~\bibnamefont{Carpenter}},
  \bibinfo{author}{\bibfnamefont{A.}~\bibnamefont{DiFranzo}},
  \bibinfo{author}{\bibfnamefont{M.}~\bibnamefont{Mulhearn}},
  \bibinfo{author}{\bibfnamefont{C.}~\bibnamefont{Shimmin}},
  \bibinfo{author}{\bibfnamefont{S.}~\bibnamefont{Tulin}}, \bibnamefont{et~al.}
  (\bibinfo{year}{2013}), \eprint{1312.2592}.

\bibitem[{\citenamefont{D'Eramo}(2007)}]{D'Eramo:2007ga}
\bibinfo{author}{\bibfnamefont{F.}~\bibnamefont{D'Eramo}},
  \bibinfo{journal}{Phys.Rev.} \textbf{\bibinfo{volume}{D76}},
  \bibinfo{pages}{083522} (\bibinfo{year}{2007}), \eprint{0705.4493}.

\bibitem[{\citenamefont{Pospelov and Ritz}(2011)}]{Pospelov:2011yp}
\bibinfo{author}{\bibfnamefont{M.}~\bibnamefont{Pospelov}} \bibnamefont{and}
  \bibinfo{author}{\bibfnamefont{A.}~\bibnamefont{Ritz}},
  \bibinfo{journal}{Phys.Rev.} \textbf{\bibinfo{volume}{D84}},
  \bibinfo{pages}{113001} (\bibinfo{year}{2011}), \eprint{1109.4872}.

\bibitem[{\citenamefont{Bai et~al.}(2012)\citenamefont{Bai, Draper, and
  Shelton}}]{Bai:2011wz}
\bibinfo{author}{\bibfnamefont{Y.}~\bibnamefont{Bai}},
  \bibinfo{author}{\bibfnamefont{P.}~\bibnamefont{Draper}}, \bibnamefont{and}
  \bibinfo{author}{\bibfnamefont{J.}~\bibnamefont{Shelton}},
  \bibinfo{journal}{JHEP} \textbf{\bibinfo{volume}{1207}}, \bibinfo{pages}{192}
  (\bibinfo{year}{2012}), \eprint{1112.4496}.

\bibitem[{\citenamefont{Greljo et~al.}(2013)\citenamefont{Greljo, Julio,
  Kamenik, Smith, and Zupan}}]{Greljo:2013wja}
\bibinfo{author}{\bibfnamefont{A.}~\bibnamefont{Greljo}},
  \bibinfo{author}{\bibfnamefont{J.}~\bibnamefont{Julio}},
  \bibinfo{author}{\bibfnamefont{J.~F.} \bibnamefont{Kamenik}},
  \bibinfo{author}{\bibfnamefont{C.}~\bibnamefont{Smith}}, \bibnamefont{and}
  \bibinfo{author}{\bibfnamefont{J.}~\bibnamefont{Zupan}},
  \bibinfo{journal}{JHEP} \textbf{\bibinfo{volume}{1311}}, \bibinfo{pages}{190}
  (\bibinfo{year}{2013}), \eprint{1309.3561}.

\bibitem[{\citenamefont{Chetyrkin et~al.}(1997)\citenamefont{Chetyrkin, Kniehl,
  and Steinhauser}}]{Chetyrkin:1996ke}
\bibinfo{author}{\bibfnamefont{K.}~\bibnamefont{Chetyrkin}},
  \bibinfo{author}{\bibfnamefont{B.~A.} \bibnamefont{Kniehl}},
  \bibnamefont{and}
  \bibinfo{author}{\bibfnamefont{M.}~\bibnamefont{Steinhauser}},
  \bibinfo{journal}{Nucl.Phys.} \textbf{\bibinfo{volume}{B490}},
  \bibinfo{pages}{19} (\bibinfo{year}{1997}), \eprint{hep-ph/9701277}.

\bibitem[{\citenamefont{Zhou et~al.}(2013)\citenamefont{Zhou, Berge, and
  Whiteson}}]{Zhou:2013fla}
\bibinfo{author}{\bibfnamefont{N.}~\bibnamefont{Zhou}},
  \bibinfo{author}{\bibfnamefont{D.}~\bibnamefont{Berge}}, \bibnamefont{and}
  \bibinfo{author}{\bibfnamefont{D.}~\bibnamefont{Whiteson}}
  (\bibinfo{year}{2013}), \eprint{1302.3619}.

\bibitem[{\citenamefont{Shoemaker}(2013)}]{Shoemaker:2013tda}
\bibinfo{author}{\bibfnamefont{I.~M.} \bibnamefont{Shoemaker}},
  \bibinfo{journal}{Phys.Dark Univ.} \textbf{\bibinfo{volume}{2}},
  \bibinfo{pages}{157} (\bibinfo{year}{2013}), \eprint{1305.1936}.

\bibitem[{\citenamefont{Cornell et~al.}(2013)\citenamefont{Cornell, Profumo,
  and Shepherd}}]{Cornell:2013rza}
\bibinfo{author}{\bibfnamefont{J.~M.} \bibnamefont{Cornell}},
  \bibinfo{author}{\bibfnamefont{S.}~\bibnamefont{Profumo}}, \bibnamefont{and}
  \bibinfo{author}{\bibfnamefont{W.}~\bibnamefont{Shepherd}},
  \bibinfo{journal}{Phys.Rev.} \textbf{\bibinfo{volume}{D88}},
  \bibinfo{pages}{015027} (\bibinfo{year}{2013}), \eprint{1305.4676}.

\bibitem[{\citenamefont{Agrawal et~al.}(2012)\citenamefont{Agrawal, Blanchet,
  Chacko, and Kilic}}]{Agrawal:2011ze}
\bibinfo{author}{\bibfnamefont{P.}~\bibnamefont{Agrawal}},
  \bibinfo{author}{\bibfnamefont{S.}~\bibnamefont{Blanchet}},
  \bibinfo{author}{\bibfnamefont{Z.}~\bibnamefont{Chacko}}, \bibnamefont{and}
  \bibinfo{author}{\bibfnamefont{C.}~\bibnamefont{Kilic}},
  \bibinfo{journal}{Phys.Rev.} \textbf{\bibinfo{volume}{D86}},
  \bibinfo{pages}{055002} (\bibinfo{year}{2012}), \eprint{1109.3516}.

\bibitem[{\citenamefont{Aprile et~al.}(2012)}]{Aprile:2012nq}
\bibinfo{author}{\bibfnamefont{E.}~\bibnamefont{Aprile}} \bibnamefont{et~al.}
  (\bibinfo{collaboration}{XENON100 Collaboration}),
  \bibinfo{journal}{Phys.Rev.Lett.} \textbf{\bibinfo{volume}{109}},
  \bibinfo{pages}{181301} (\bibinfo{year}{2012}), \eprint{1207.5988}.

\bibitem[{\citenamefont{Akerib et~al.}(2013)}]{Akerib:2013tjd}
\bibinfo{author}{\bibfnamefont{D.}~\bibnamefont{Akerib}} \bibnamefont{et~al.}
  (\bibinfo{collaboration}{LUX Collaboration}) (\bibinfo{year}{2013}),
  \eprint{1310.8214}.

\bibitem[{\citenamefont{Saab}(SuperCDMS Collaboration, Talk at SSI
  2012)}]{SCDMS}
\bibinfo{author}{\bibfnamefont{T.}~\bibnamefont{Saab}}
  (\bibinfo{year}{SuperCDMS Collaboration, Talk at SSI 2012}).

\bibitem[{\citenamefont{Aprile}(2012)}]{Aprile:2012zx}
\bibinfo{author}{\bibfnamefont{E.}~\bibnamefont{Aprile}}
  (\bibinfo{collaboration}{XENON1T collaboration}) (\bibinfo{year}{2012}),
  \eprint{1206.6288}.

\bibitem[{\citenamefont{Hewett and Rizzo}(1989)}]{Hewett:1988xc}
\bibinfo{author}{\bibfnamefont{J.~L.} \bibnamefont{Hewett}} \bibnamefont{and}
  \bibinfo{author}{\bibfnamefont{T.~G.} \bibnamefont{Rizzo}},
  \bibinfo{journal}{Phys.Rept.} \textbf{\bibinfo{volume}{183}},
  \bibinfo{pages}{193} (\bibinfo{year}{1989}).

\bibitem[{\citenamefont{Langacker}(2009)}]{Langacker:2008yv}
\bibinfo{author}{\bibfnamefont{P.}~\bibnamefont{Langacker}},
  \bibinfo{journal}{Rev.Mod.Phys.} \textbf{\bibinfo{volume}{81}},
  \bibinfo{pages}{1199} (\bibinfo{year}{2009}), \eprint{0801.1345}.

\bibitem[{\citenamefont{Arzt}(1995)}]{Arzt:1993gz}
\bibinfo{author}{\bibfnamefont{C.}~\bibnamefont{Arzt}},
  \bibinfo{journal}{Phys.Lett.} \textbf{\bibinfo{volume}{B342}},
  \bibinfo{pages}{189} (\bibinfo{year}{1995}), \eprint{hep-ph/9304230}.

\end{thebibliography}

\end{document}